\documentclass[aps,prd,twocolumn,showpacs,amsmath,nofootinbib]{revtex4-1}

\usepackage{color}
\newcommand{\beqa}{\begin{eqnarray}}
\newcommand{\eeqa}{\end{eqnarray}}
\usepackage{float}
\usepackage{epsfig}
\usepackage{graphicx}
\usepackage{latexsym}
\usepackage{amsmath}
\usepackage{hyperref}
\usepackage{mathrsfs}
\usepackage{dcolumn}
\usepackage{bm}%

\begin{document}

\title{Cosmological constraints on the big bang quantum cosmology model}

\author{Yicheng Wang$^1$}\email{939507273@qq.com}
\author{Yupeng Yang$^{1}$}\email{Corresponding author:  ypyang@aliyun.com}
\author{Xinyi Dai $^1$}\email{1584377516@qq.com}
\author{Shuangxi Yi $^1$}\email{yisx2015@qfnu.edu.cn}
\author{Yankun Qu $^1$}\email{quyk@qfnu.edu.cn}
\author{Fayin Wang $^{2,3}$}\email{fayinwang@nju.edu.cn}
\affiliation{$^1$School of Physics and Physical Engineering, Qufu Normal University, Qufu, Shandong, 273165, China \\
           $^2$ School of Astronomy and Space Science, Nanjing University, Nanjing 210023, China\\
           $^3$ Key Laboratory of Modern Astronomy and Astrophysics (Nanjing University) Ministry of Education,  China\\ }

\begin{abstract}
The big bang quantum cosmology model introduces the trace $J$ of the Schouten tensor as a form of dynamic dark energy. Together with cold dark matter, these components form
the so-called $J$CDM cosmology model, proposed by M.H.P.M. van Putten (J. High Energy Astrophys., 45, 2025, 194), which offers a potential resolution to the Hubble tension.
We derive the constraints on the $J$CDM cosmology model, utilizing early- and late-time cosmological data including cosmic microwave background (CMB), baryon acoustic oscillations (BAO) released by the Dark Energy Spectroscopic Instrument (DESI), cosmic chronometers (CC), and type Ia supernovae (SNIa). For a flat universe, the $J$CDM model yields \( H_0 = 66.95 \pm 0.51 \, \rm{km~s^{-1}~Mpc^{-1}} \) and \( \Omega_m = 0.3419 \pm 0.0065 \), results that are consistent with early-universe observations but exhibit a higher \( \Omega_m \) compared to the $\Lambda$CDM model. In the case of a non-flat universe, $J$CDM favors a slightly curved geometry with \( \Omega_k = 0.0154 \pm 0.0027 \), leading to \( H_0 = 69.13 \pm 0.56 \, \rm {km~s^{-1}~Mpc^{-1}} \) and \( \Omega_m = 0.3477 \pm 0.0074 \). The increase in \( H_0 \) in the non-flat scenario suggests a geometric degeneracy between spatial curvature and \( H_0 \). We also investigate the internal inconsistencies present in DESI data and evaluate their impacts on cosmological parameter constraints. 
Our analysis shows that while the $J$CDM model, which is constructed from first principles without free parameters beyond those of $\Lambda$CDM, agrees excellently with late-time cosmology, it struggles to simultaneously match early-universe observations in a fully self-consistent manner.

\end{abstract}

\maketitle

\section{Introduction}
The standard cosmological model $\Lambda$CDM has achieved remarkable success in describing the evolution of the Universe, particularly in matching observations of the cosmic microwave background (CMB)~\cite{2021Cosmological,2018Planck}. However, while it remains qualitatively consistent, its predictions are quantitatively less definitive for large-scale structure~\cite{Tinker_2011} and late-time cosmic acceleration~\cite{SupernovaSearchTeam:1998fmf,SupernovaCosmologyProject:1998vns}.
Moreover, several persistent tensions challenge its completeness, with the most notable being the Hubble tension, the discrepancy between the early-universe inference of the Hubble constant \( H_0 = 67.36\pm 0.54\, \text{km s}^{-1} \text{Mpc}^{-1} \) from the Planck satellite \cite{2018Planck} and the late-universe measurement of \( H_0 =73.04\pm {1.04} \, \text{km s}^{-1} \text{Mpc}^{-1} \) via the local distance ladder~\cite{2021A,Riess_2009,Breuval_2024,2020The,Verde_2019} 
(see Table~\ref{table:h0}). Furthermore, the discrepancy between the predicted and observed amplitudes of matter clustering further underscores the potential limitations of the $\Lambda$CDM model assumptions, particularly its static dark energy (DE) component, namely the cosmological constant $\Lambda$~\cite{Bull:2015stt}. Current efforts to resolve these tensions encompass a range of approaches~\cite{DiValentino:2021izs,Hu:2023jqc,Dainotti:2021pqg,Yang:2025oax,Li:2025muv,CosmoVerseNetwork:2025alb}: the interaction between dark matter and dark energy~\cite{Wang_2016,Naidoo_2024,PhysRevD.94.043518,PhysRevD.96.043503,PhysRevD.96.103511,PhysRevD.98.123527,Zhang:2025dwu,Li:2024qso,Li:2025owk,Li:2026xaz}, early dark energy model~\cite{Kamionkowski:2022pkx,Poulin_2019}, vacuum decay model~\cite{Overduin_1998,Arbab_2002,Shapiro_2002,Shapiro_2003,Espa_a_Bonet_2004,Koussour_2024,Azri_2016,Szyd_owski_2015,PhysRevD.105.063532,Azri_2012,Yang:2025vnm}, dynamic dark energy models~\cite{COPELAND_2006,G_mez_Valent_2015,Sol__2015,Sol__2017,Sol_Peracaula_2018,Sol_Peracaula_2018_1,Sol__2017_1,ray2007acceleratinguniversedynamiccosmological,Alam:2025epg,RoyChoudhury:2025iis,Wu:2025vfs} (e.g., \( w(z) \)CDM), modified gravity theories~\cite{Perivolaropoulos_2022}, and the extended models that incorporate additional relativistic particle species or non-zero spatial curvature~\cite{wong2023fastgravitationalwaveparameter,Wolf:2025acj,Zhou:2025nkb}. However, many proposed solutions introduce fine-tuning issues or conflict with early-universe observational data. A critical challenge lies in reconciling the dynamics of late-time dark energy with the precise angular scales of the cosmic microwave background (CMB) acoustic peaks, which impose stringent constraints on early universe physics~\cite{Poulin_2019}.

The $J$CDM model has emerged as a novel framework rooted in quantum cosmology to address the aforementioned contradictions by deriving from first principles that dark energy is a relic of the big bang. Unlike $\Lambda$CDM, $J$CDM posits that the dynamics of dark energy originate from the trace \( J \) of the Schouten tensor~\cite{van_Putten_2025}, a geometric quantity associated with conformal symmetry. This approach stems from the infrared-consistent coupling of the bare cosmological constant with spacetime, adhering to the Bekenstein entropy bound. As a result, the dark energy density scales as $\rho_{\Lambda} \propto J $, which varies with the Hubble parameter $H(z)$.
The predictive power of $J$CDM hinges on its analytical solution for $H(z)$, which is calibrated against the baryon acoustic oscillation (BAO) scale observed in the cosmic microwave background (CMB)~\cite{vanputten2024hubbleparameterlocaldistance}. The model introduces a scaling relation
$H_0 = \sqrt{6/5} \, H_0^{\Lambda}$, where $H_0^{\Lambda}$ is the Hubble constant in $\Lambda$CDM cosmology.
Through a simple estimation, \( H_0 \) can be approximated to $74~\rm km s^{-1} Mpc^{-1}$, and fitting to $H(z)$ data gives 
$H_{0}=74.9\pm{2.60} ~\rm km s^{-1} Mpc^{-1}$~\cite{van_Putten_2025}, as presented in Table~\ref{table:h0}. 
Simultaneously, the matter density
\( \Omega_{m} \) is reduced to about 0.26~\cite{van_Putten_2025}, which is consistent with constraints from the local distance ladder. Crucially, $J$CDM preserves the CMB acoustic horizon \( r_* \) and angular scale \( \theta_* \), ensuring compatibility with Planck data. The equation of state
for dynamic dark energy, given by \( w = (2q - 1)/(1 - q) \) (where \( q \) is the deceleration parameter), predicts a value of \( q_0 \simeq -1 \).
This prediction significantly deviates from the \( q_0 \simeq -0.5 \) anticipated by the $\Lambda$CDM model\footnote{Note that the value of 
$q_0$ reported in Ref.~\cite{2021A} is 
model-dependent, as it is derived under $\Lambda$CDM assumptions with $j_0=1$~\cite{Riess:2022oxy} (v1 version).}, thereby providing a testable
characteristic for low-redshift surveys.

The authors of~\cite{van_Putten_2025} utilized a streamlined methodology to estimate the values of cosmological parameters,
yet they did not account for the correlations among these parameters. In this work, we employ a comprehensive array of
cosmological datasets, baryon acoustic oscillations (BAO) data from the Dark Energy Spectroscopic Instrument (DESI DR1)
~\footnote{A description of the acronyms used in the main text can be found in Table~\ref{table:acronym}.},
cosmic microwave background (CMB) measurements, encompassing cosmic chronometers (CC), and observations of type Ia supernovae (SNIa).
Furthermore, we employ the Markov Chain Monte Carlo (MCMC) method to rigorously constrain the parameters of the $J$CDM model.
Our results indicate the potential of the $J$CDM model to harmonize the apparent discrepancies between early-time and late-time
cosmological observations.

It is important to note that the current analysis relies on DESI BAO data. As highlighted, e.g., in Refs.~\cite{2024confirm,chaudhary2025}, this dataset may exhibit internal inconsistencies, particularly an anomalously high matter density reported from the luminous red galaxy (LRG) sample at the effective redshift \( z_{\text{eff}} = 0.51 \), which is in approximately $2\sigma$ tension with the Planck measurements and supernova constraints, respectively. Although DESI reports consistency with earlier SDSS results, these discrepancies indicate that cosmological parameter constraints derived from this dataset should be interpreted with caution, as reconstruction of the BAO feature in the matter power spectrum corrects for cosmic evolution using a fiducial $\Lambda$CDM cosmology and may therefore introduce finite, not yet quantified, model dependence. 
In this work, we will investigate the impact of this sample on final constraints of $J$CDM cosmological model.

The structure of this paper is as follows: In Sec.~\ref{sec:basic}, we provide an overview of the fundamental elements of the $J$CDM model;
Sec.~\ref{sec2} introduces the datasets used in this study, along with an explanation of the underlying principles and
the methodologies employed for data processing. The results and relevant discussions are given in Sec.~\ref{sec:III}. 
Finally, Sec.~\ref{sec:conclusions} summarizes the key findings and presents our conclusions.

\section{The basic properties of $J$CDM cosmology model}
\label{sec:basic}
In the framework of post-big bang quantum cosmology, it is hypothesized that time-translation symmetry is violated on the Hubble time scale, leading to the presence of vacuum energy in the form of residual thermal energy\footnote{Note: More precisely, it is a geometric vacuum energy residing in the off-shell thermal structure of the Hubble horizon geometry.}. This dynamic dark energy component is characterized by the trace $J$ of the Schouten tensor~\cite{vanputten2024hubbleparameterlocaldistance,van_Putten_2025},
\beqa
J=\frac{1}{6} R
\eeqa
where $R$ is the scalar curvature. Utilizing the path integral approach, a global phase gauge is introduced to account for
and absorb the effects of quantum fluctuations, thereby enabling the derivation of the relationship between dark energy
and the Hubble parameter~\cite{1955Tensor,van_Putten_2015},
\beqa
\Lambda =J
\eeqa
with
\beqa
J\equiv (1-q)\frac{H^{2}}{c^{2}}
\eeqa
here, $q$ denotes the deceleration parameter, and $J$ differs from the cosmological constant $\Lambda$ in $\Lambda$CDM cosmology.
Specifically, $J$ is a dynamic quantity that evolves with the expansion of the Universe.

In the $J$CDM cosmology model, by modifying the Friedmann equation, an analytical solution of the Hubble rate has been derived in~\cite{VANPUTTEN2021136737}:
\beqa
h(z)=\frac{\sqrt{1+\frac{3}{2}\Omega_{k}Z_{4}(z)+\frac{6}{5}\Omega_{m}Z_{5}(z)+\Omega_{r}Z_{6}(z)    } }{1+z}
\eeqa
here, $Z_{n}(z) = (1+z)^{n} - 1$ , \(\Omega_{m}\), \(\Omega_{r}\), and \(\Omega_{k}\) represent the dimensionless density of matter,
the radiation density at \(z = 0\), and the curvature density, respectively.
\(\Omega_{k} < 0\) (\(> 0\)) corresponds to negative (positive) curvature.

\begin{table}[htb]
\caption{Baseline values of $H_0$ in the $\Lambda$CDM and $J$CDM cosmological models. “Planck/CMB” denotes the result derived from Planck 2018 CMB data~\cite{2018Planck}, “SHOES/LDL” is the model-independent measurement from the SHOES local distance ladder~\cite{2021A}, and “CC” indicates the model-independent estimate based on cosmic chronometer data~\cite{Farooq:2016zwm}. “$\rm LDL_{cubic}$” represents the cubic polynomial fit to the $H(z)$ data~\cite{van_Putten_2025,van_Putten_2017}. The value of $H_0$ for the $\Lambda$CDM model using cosmic chronometer data is labeled as $\Lambda\rm CDM^{CC}$~\cite{van_Putten_2017}. The $J$CDM model prediction for $H_0$ obtained from fitting the $H(z)$ data is labeled as “$ J$$\rm CDM^{LDL}$”~\cite{van_Putten_2025}.}

\begin{ruledtabular}
\begin{tabular}{ll}


Model/dataset &  $H_0[\rm km~s^{-1}~Mpc^{-1}]$ \\

\noalign{\smallskip}\hline\noalign{\smallskip}

$\rm SHOES/LDL$   &$73.04\pm{1.04}$    \\

$\rm CC$   &$73.24\pm{1.74}$   \\

$\rm LDL_{\rm cubic}$   &$74.44\pm{4.9}$  \\

$\rm \Lambda CDM^{Planck/CMB}$   &$67.36\pm{0.54}$  \\ 

$\rm \Lambda CDM^{CC}$   &$66.8\pm{1.9}$  \\ 

$J{\rm CDM^{LDL}}$  &$74.9\pm{2.60}$   \\
\end{tabular}
\end{ruledtabular}
\label{table:h0}
\end{table}

\begin{table}[htb]
\renewcommand{\arraystretch}{0.6} 
\caption{Description of acronyms appearing in the main text}

\begin{ruledtabular}
\begin{tabular}{l p{5cm}}
\noalign{\smallskip}

Acronym & Description \\

\noalign{\smallskip}\hline\noalign{\smallskip}

Planck CMB   & the Cosmic Microwave Background detected by Planck satellite experiment  \\
\\
SH0ES  & for Supernova H0 for the Equation of State\\
\\
BAO   & Baryon Acoustic Oscillations   \\
\\
DESI DR1   & Data Release 1 of Baryon Acoustic Oscillations data from first 13 months of the Dark Energy Spectroscopic Instrument main survey \\
\\
CC   & Cosmic Chronometers  \\ 
\\
SNIa  & type Ia Supernova   \\
\\
BGS  & Bright Galaxy Sample (for BAO data)  \\
\\
LRG  & Luminous Red Galaxy Sample (for BAO data)  \\
\\
ELG  & Emission Line Galaxy Sample (for BAO data)  \\
\\
QSO  & Quasar Sample (for BAO data)   \\
\\
LFS  & Lyman-$\alpha$ Forest Sample (for BAO data)   \\
\\
LDL  & Local Distance Ladder   \\

\noalign{\smallskip}
\end{tabular}
\end{ruledtabular}
\label{table:acronym}
\end{table}

\section{Data used for analysis}
\label{sec2}

\subsection{Baryon acoustic oscillations}

In this study, we employ the Baryon Acoustic Oscillation (BAO) data released firstly from the Dark Energy Spectroscopic Instrument (DESI DR1)~\cite{desicollaboration2024desi2024vicosmological}. The sample includes the following sub-samples:
Bright Galaxy Sample (BGS, $z_{\rm eff}=0.30$),
Luminous Red Galaxy Sample (LRG, $z_{\rm eff}$=0.51 and 0.71),
Emission Line Galaxy Sample (ELG, $z_{\rm eff}=1.32$),
the combined LRG and ELG sample (LRG+ELG, $z_{\rm eff}=0.93$),
Quasar Sample (QSO, $z_{\rm eff}=1.49$),
and Lyman-$\alpha$ Forest Sample (Ly$\alpha$, $z_{\rm eff}=2.33$).

Within a homogeneous and isotropic cosmological framework,
the transverse comoving distance is
\beqa
D_{M}(z)=\frac{c}{H_{0}\sqrt{|\Omega_{k}|} }{\rm sinn} \left [ \sqrt{|\Omega_{k}|} \int_{0}^{z}  \frac{dz^{' }}{H(z^{'})/H_{0}} \right ]
\eeqa
here, $c$ denotes the speed of light, and $H(z)$ is the Hubble rate. ${\rm sinn}(x) = {\rm sinh}(x), x, {\rm sin}(x)$ corresponds an open, flat and closed universe, respectively. 
The equivalent distance variable $D_{H}(z)$ is defined as $D_{H}(z) = c / H(z)$~\cite{DiValentino:2021izs,Hu:2023jqc,Yang:2025oax}. The expression for the angular average distance
$D_{\rm V}$ is constructed by incorporating the relationship between $D_{\rm M}$ and $D_{\rm H}$, which can be written as~\cite{Eisenstein_2005}
\beqa
D_{V}(z) = [z D_{M}^2(z) D_{H}(z)]^{1/3}.
\eeqa

For the standardized distance parameters \(D_M/r_d\) and \(D_H/r_d\) from the samples of the Bright Red Galaxy (LRG, with effective redshifts \(z_{\text{eff}} = 0.51\) and \(0.71\)), the LRG + Emission Line Galaxy (ELG) combined sample, the ELG sample, and the Lyman-alpha Forest Quasar (Ly\(\alpha\) QSO) sample,
we adopt the acoustic horizon $r_{d}$ given in~\cite{Brieden_2023}, 

\beqa
r_{d}=\frac{147.05}{\rm Mpc} \left(\frac{\Omega _{m}h^2}{0.1432} \right)^{-0.23}\left(\frac{N_{\rm eff}}{3.04} \right)^{-0.1}\left(\frac{\Omega _{b}h^2}{0.02236} \right)^{-0.13}
\label{eq:rd}
\eeqa

We fix the effective number of relativistic degrees of freedom $N_{\rm eff} = 3.04$ to maintain a standardized radiation content in the early universe, while allowing the matter density parameters to vary freely in accordance with observational data, thereby improving the fit for the Hubble constant.

The chi-square statistics can be expressed as~\cite{Kamionkowski:2022pkx}
\beqa
{\chi ^{2} _{1}}=\sum_{i}^{} \bigtriangleup D_{i}^{T}{\rm Cov}_{\rm BAO}^{-1}\bigtriangleup D_{i},
\eeqa
where
\beqa
D_{i}=\begin{bmatrix}D_{M}/r_{d}\\D_{H}/r_{d}\end{bmatrix}
\eeqa
and
$\Delta D_{i}=D_{i}^{\rm th}-D_{i}^{\rm data}$.
For the standardized volume distance parameter \(D_V/r_d\) of the bright galaxy sample (BGS, \(z_{\text{eff}}=0.30\)) and the quasar sample (QSO, \(z_{\text{eff}}=1.49\)), the chi-square statistic can be expressed as

\beqa
\chi^2 _{2}= \sum_{k} \left( \frac{D_V^{th}/r_d - D_V^{\text{obs}}/r_d}{\sigma_{D_V}} \right)^2.
\eeqa

Therefore, for the DESI BAO data, the $\chi^{2}$ can be written as
\beqa
\chi^2 _{\rm DESI}=\chi^2 _{1}+\chi^2 _{2}
\eeqa

As mentioned above, the DESI collaboration has provided measurements of the BAO scale across multiple redshift bins in the form of the ratios \(D_M(z)/r_d\) and \(D_H(z)/r_d\). A fundamental characteristic of these measurements is that they constrain only relative distance scales, resulting in a well known degeneracy between the absolute distance scale, specifically the Hubble constant \(H_0\), and the sound horizon at the drag epoch, \(r_d\). To break this degeneracy and enable robust inference of key cosmological parameters such as \(H_0\) and \(\Omega_m\), an external calibration of \(r_d\) is required. 
As shown in Refs.~\cite{2024confirm,chaudhary2025}, the DESI collaboration employed a prior of \(r_d = 147.09\pm 0.26~\text{Mpc}\) derived from the observations of the cosmic microwave background (CMB) by the Planck satellite for cosmological calibration. In this work, we also use this prior setting to derive cosmological parameter constraints and perform comparative analyses to evaluate its implications.

Moreover, the DESI DR1 data exhibit notable internal inconsistencies across different redshift probes. Specifically, in the LRG sample at an effective redshift of $z_{\text{eff}} = 0.51$, the inferred matter density parameter is elevated (\(\Omega_{m}\) $=$ 0.67)~\cite{2024confirm,chaudhary2025}, deviating from the value derived from the PantheonPlus supernova sample by approximately 2$\sigma$~\cite{tdcosmo}. Furthermore, the DESI data display unphysical and pronounced fluctuations in \(\Omega_{m}\) across various redshift bins, a feature that may arise from unresolved systematic uncertainties or statistical noise. To address this issue, we will conduct an independent analysis of the $z_{\text{eff}} = 0.51$ LRG sample in following sections, investigating the impacts on the final constraints with and without the inclusion of this sample.

Although the DESI DR1 data exhibit some degree of internal inconsistency, their strong capability in constraining cosmological models has led to their wide application. Results derived from DESI DR1 data can be compared with those from other existing studies. More importantly, 
DESI DR1 data may provide a valuable reference for cosmological research focused on the late-time expansion history. Naturally, caution is required when using these data, and careful assessment is necessary in analyzing and interpreting the results obtained from them.

\subsection{Cosmic microwave background} 
Typically, constraints on cosmological parameters from Cosmic Microwave Background (CMB) data are derived from measurements of the angular power spectrum. However, for models that deviate from $\Lambda$CDM during cosmic epochs where the overall shape of the power spectrum remains largely unaffected, CMB distance priors can provide an effective alternative for constraining cosmological parameters~\cite{Chen_2019,Wang:2007mza,Zhai:2019nad}. In addition to providing constraints equivalent to those from the full CMB angular power spectrum, the CMB distance prior yields results comparable to the more time-consuming power-spectrum analysis, while requiring substantially less computational resources and time. This approach has been widely used in previous studies (see, e.g., Refs.~\cite{Yang:2025vnm,Yang:2025oax,8ync-vrtz,Zhai:2018vmm,Li:2024hrv,Jia:2025prq,Rezaei:2024vtg,Sohail:2024oki}). 
In this work, we therefore employ the CMB distance priors, which include the sound horizon encoded in the acoustic scale $l_a$, the shift parameter $R$, and the baryon density $\Omega_b h^2$, to derive robust constraints on the cosmological parameters~\cite{Liu_2019, Xu_2016, Komatsu_2009, Yang:2025oax}. It should be noted that, while the distance prior is used here to obtain final constraints on the relevant parameters, more accurate constraints would in principle be achieved from a full analysis of the CMB angular power spectrum data. 

The acoustic scale and shift parameter are defined as 

\beqa
&&l_{a}\equiv  (1+z_{*})\frac{\pi D_{A}(z_{*})}{r_{s}(z_{*})}\\
&&R\equiv \sqrt{\Omega _{m}H_{0}^2} (1+z_{*})D_{A}(z_{*})
\eeqa
here, $z_{*}$ represents the redshift at the epoch of photon decoupling, and \(r_{s}\) is the comoving sound horizon. We adopt an approximate form for \(z_{*}\) as follows~\cite{Hu_1996}:
\beqa
z_{*}=1048[1+0.00124(\Omega_{b}h^{2} )^{-0.738}][1+g_{1}(\Omega_{m}h^{2})^{g_{2}}]
\eeqa
with
\beqa
g_{1}&=&\frac{0.0783(\Omega_{b}h^{2})^{-0.238}}{1+39.5(\Omega_{b}h^{2})^{0.763}}\\
g_{2}&=&\frac{0.56}{1+21.1(\Omega_{b}h^{2})^{1.81}}
\eeqa

$r_{s}$ can be expressed as
\beqa
r_{s}(z)=\frac{c}{H_{0}} \int_{0}^{1/(1+z)} \frac{da}{a^{2}h(a)\sqrt{3(1+\frac{3\Omega_{b}h^{2}}{4\Omega_{\gamma }h^{2}}a )}},
\eeqa
where $a=1/(1+z)$ and
\beqa
\frac{3}{4\Omega_{\gamma }h^{2}}=31500(T_{\rm CMB}/2.7K)^{-4}
\eeqa
with $T_{\rm CMB}$=2.7255K. The angular diameter distance $D_A$ can be written as
\beqa
D_{A}(z)=\frac{c}{(1+z)H_{0}\sqrt{|\Omega_{k}|} }{\rm sinn}\left [ \sqrt{|\Omega_{k}|} \int_{0}^{z}  \frac{dz^{' }}{H(z^{'})/H_{0}} \right ]
\eeqa

For CMB data, the $\chi^2$ can be written as
\beqa
\chi ^{2}_{\rm CMB}=\Delta X^{T}{\rm Cov}^{-1}_{\rm CMB}\Delta X,
\eeqa
here, $\Delta X$=$X$-$X^{\rm obs}$, and \( X^{\text{obs}} \) is formulated as a composite observational vector comprising three key parameters: the shift parameter \( R \), the acoustic scale \( l_a \), and the baryon density parameter \( \Omega_b h^2 \). The numerical values of the observation vector \( X^{\text{obs}} \) and the inverse covariance matrix \( \mathbf{\rm Cov}^{-1}_{\rm CMB} \) utilized in this study are sourced from the Planck 2018 as given in, e.g, Ref.~\cite{Chen_2019}.
\subsection{Cosmic chronometers}
We have acquired data regarding the rate of cosmic expansion by utilizing cosmic chronometers, which are based on the relative ages of galaxies, their peak masses, and passively evolving galaxies~\cite{Jimenez_2002}. The Hubble parameter, as determined by the Cosmic Chronometer (CC) method~\cite{Jimenez_2002,Moresco_2022}, is expressed as
\beqa
H(z)=-\frac{1}{1+z} \frac{\Delta z}{\Delta t}
\eeqa

For CC data, the $\chi^2$ can be written as,
\beqa
\chi _{\rm CC}^{2}=\sum_{i}^{} \frac{\left[H(z_{i})-H_{\rm obs}(z_{i})\right]^{2}}{\sigma ^{2}}
\eeqa
and we utilize 31 cosmic chronometer (CC) data points derived from a variety of scholarly sources~\cite{Moresco_2016,Pal_2024,Moresco_2015,Ratsimbazafy_2017,Zhang_2014}.
\subsection{Type Ia supernova}
For the observation of supernovae, luminosity distance  $d_l$ is an important parameter, which can be expressed as~\cite{Gong_2007,Yang:2025oax}:
\beqa
d_{l}(z_{i})=\frac{1+z}{\sqrt{\left | \Omega_{k} \right |}}{\rm sinn}\left \{ \sqrt{\left | \Omega_{k} \right |} \int_{0}^{z} \frac{cdz^{'}}{H_{0} h(z^{'})}\right \}
\eeqa
where ${\rm sinn}(x) = {\rm sinh}(x), x, {\rm sin}(x)$ corresponds an open, flat and closed universe, respectively.
The theoretical value of the distance modulus can be calculated as
\beqa
\mu_{\rm th}(z_{i} )=5{\rm log_{10}}D_{L}(z_{i})+25
\eeqa
By marginalizing over the Hubble constant as detailed in the literature, we are able to deduce that  \cite{PhysRevD.72.123519}
\beqa
&&\chi _{\rm SN}^{2}(\theta ) =A(\theta )-\frac{B^{2}(\theta )}{C}
\eeqa
where
\beqa
&&A(\theta )=\sum_{i=1}^{N}\frac{(\mu _{\rm obs}(z_{i})-5{\rm log_{10}}(D_{L}(z_{i}))  )^{2}}{\sigma _{i}^{2}} \\
&&B(\theta )=\sum_{i=1}^{N}\frac{\mu _{\rm obs}(z_{i})-5{\rm log_{10}}(D_{L}(z_{i})  )}{\sigma _{i}^{2}} \\
&&C(\theta )=\sum_{i=1}^{N}\frac{1}{\sigma _{i}^{2}} \\
&&D_{L}(z_{i})=\frac{H_{0}}{c} \times d_{l}(z_{i})
\eeqa
In this study, we utilize a sample of 1048 supernovae from the Pantheon compilation, spanning a redshift range of 0.01 and 2.3~\cite{Scolnic_2018}.
\section{Model analysis, results, and relevant discussions}
\label{sec:III}
We will incorporate data from various sources, including Baryon Acoustic Oscillations (BAO) measurements from the Dark Energy Spectroscopic
Instrument (DESI DR1), Cosmic Microwave Background (CMB) data, cosmic chronometers (CC), and observations of type Ia supernovae (SNIa).
By utilizing the Markov Chain Monte Carlo (MCMC) technique, our objective is to ascertain the optimal values and posterior
distributions for the $J$CDM cosmological parameters. Specifically, for the flat universe model, the parameters of interest are
\(\{\Omega_{m}, H_{0}, \Omega_{b}h^{2}\}\), while for the non-flat universe model, we consider the parameter set 
\(\{\Omega_{m}, \Omega_{k}, H_{0}, \Omega_{b}h^{2}\}\).

The analysis will be performed under both flat and non-flat cosmological frameworks to assess the consistency of
the $J$CDM model with observational constraints. The total $\chi^{2}$ can be written as
\beqa
\chi ^{2}_{\rm total}=\chi ^{2}_{\rm DESI}+\chi ^{2}_{\rm CMB}+\chi ^{2}_{\rm CC}+\chi ^{2}_{\rm SN}
\eeqa
Then the total likelihood function is given by $L\propto e^{-\chi^{2}_{\rm total}/2}$. We utilize the open-source code
\texttt{emcee} for MCMC sampling. The uniform prior distributions for the parameters are set as:
$\Omega_{k}\in(-0.4,0.4)$, $\Omega_{m}\in(0,1)$, $H_{0}\in(50,90)$, and $\Omega_{b}h^{2}\in(0,0.1)$. Additionally, we employ the open-source
package \texttt{GetDist}~\cite{lewis2019getdistpythonpackageanalysing} to analyze the MCMC chains~\cite{emcee}.
\begin{table*}[htb]
\caption{Constraints on the parameters of $J$CDM model in flat and non-flat universe, derived from different combined datasets using the acoustic horizon scale $r_{d}$ defined in Eq.~(\ref{eq:rd}) (referred to as the ``unfixed'' case).}
\label{tab_ann}
\begin{center}
\begin{ruledtabular}
\begin{tabular}{lllll}
Model/dataset & $\Omega_{m}$& $\Omega_{k}$ & $H_0[\rm km~s^{-1}~Mpc^{-1}]$ &$\Omega_{b}h^{2}$ \\

\noalign{\smallskip}\hline\noalign{\smallskip}
{\bf Flat} \\

DESI+CMB   &$0.3262\pm{0.0091}$ &$-$  &$68.19\pm{0.75}$ &$0.02388\pm{0.00013}$   \\

DESI+CMB+CC  &$0.3256^{+0.0088}_{-0.0098}$ &$-$  &$68.24\pm{0.76}$ &$0.02388\pm{0.00013}$  \\

DESI+CMB+CC+SNIa  &$0.3419\pm{0.0065}$ &$-$  &$66.95\pm{0.51}$ &$0.02378\pm{0.00012}$  \\

\noalign{\smallskip}\hline\noalign{\smallskip}
{\bf Non-flat} \\

DESI+CMB   &$0.364^{+0.017}_{-0.021}$ &$0.0197^{+0.0044}_{-0.0056}$  &$68.50\pm{0.98}$ &$0.02422\pm{0.00014}$   \\

DESI+CMB+CC  &$0.349^{+0.016}_{-0.018}$ &$0.0156\pm{0.0044}$  &$69.08\pm{0.90}$ &$0.02427\pm{0.00014}$  \\

DESI+CMB+CC+SNIa  &$0.3477\pm{0.0074}$ &$0.0154\pm{0.0027}$ &$69.13\pm{0.56} $&$0.02427\pm{0.00013}$ \\

\end{tabular}
\end{ruledtabular}
\end{center}
\label{table:cons_no_rd}
\end{table*}

\begin{table*}[htb]
\caption{Constraints on the parameters of $J$CDM model in flat and non-flat universe, derived from different combined datasets using the CMB derived acoustic horizon scale $r_{d}=147.09\pm{0.29} ~\rm Mpc$ (referred to as the ``fixed" case).}
\label{tab_ann}
\begin{center}
\begin{ruledtabular}
\begin{tabular}{lllll}
Model/dataset & $\Omega_{m}$& $\Omega_{k}$ & $H_0[\rm km~s^{-1}~Mpc^{-1}]$ &$\Omega_{b}h^{2}$ \\

\noalign{\smallskip}\hline\noalign{\smallskip}
{\bf Flat} \\

DESI+CMB   &$0.385\pm{0.0013}$ &$-$  &$63.98\pm{0.87}$ &$0.02369\pm{0.00012}$   \\

DESI+CMB+CC  &$0.385^{+0.0012}_{-0.0014}$ &$-$  &$64.00\pm{0.86}$ &$0.02369\pm{0.00013}$  \\

DESI+CMB+CC+SNIa  &$0.3640\pm{0.0073}$ &$-$  &$65.40\pm{0.52}$ &$0.02380\pm{0.00012}$  \\

\noalign{\smallskip}\hline\noalign{\smallskip}
{\bf Non-flat} \\

DESI+CMB   &$0.399\pm{0.018}$ &$0.0077\pm{0.0024}$  &$64.2\pm{1.0}$ &$0.02381\pm{0.00015}$   \\

DESI+CMB+CC  &$0.389^{+0.016}_{-0.018}$ &$0.0070^{+0.0022}_{-0.0024}$  &$64.79\pm{0.99}$ &$0.02387\pm{0.00014}$  \\

DESI+CMB+CC+SNIa  &$0.3616\pm{0.0076}$ &$0.0046\pm{0.0017}$ &$66.35\pm{0.54} $&$0.02402^{+0.00011}_{-0.00013}$ \\

\end{tabular}
\end{ruledtabular}
\end{center}
\label{table:cons_rd}
\end{table*}
\begin{table*}[htb]
\caption{Constraints on the parameters of the standard $\Lambda$CDM cosmological model and $J$CDM model in both flat and non-flat universe, 
derived from different combined datasets. Results are presented for both unfixed and fixed $r_{d}$ (denoted by superscripts $r_d$ for fixed case, 
e.g., $\Lambda$${\rm CDM}^{r_d}$ 
and ${\rm JCDM}^{r_d}$). 
The constraints including and excluding the DESI BAO data at $z=0.51$ are also provided for investigating the impact of this data on final results. }
\label{tab_ann}
\begin{center}
\begin{ruledtabular}
\begin{tabular}{lllllccc}
Model/dataset & $\Omega_{m}$& $\Omega_{k}$ & $H_0[\rm km~s^{-1}~Mpc^{-1}]$ &$\Omega_{b}h^{2}$\\

\noalign{\smallskip}\hline\noalign{\smallskip}
{\bf Flat} \\

$\Lambda$CDM$^{r_{d}}$   &$0.2998^{+0.0045}_{-0.0056}$ &$-$  &$68.58^{+0.42}_{-0.36}$ &$0.02261\pm{0.00013}$   \\

$\Lambda$CDM$^{r_{d}}$(without $z=0.51$)  &$0.3019\pm{0.0051}$ &$-$  &$68.42\pm{0.39}$ &$0.02258\pm{0.00013}$  \\

$\Lambda$CDM   &$0.3001\pm{0.0255}$ &$-$  &$68.53\pm{0.43}$ &$0.02254\pm{0.00013}$   \\

$\Lambda$CDM(without $z=0.51$)  &$0.3016\pm{0.0055}$ &$-$  &$68.43\pm{0.44}$ &$0.02255\pm{0.00014}$  \\

$J$CDM$^{r_{d}}$  &$0.3640\pm{0.0073}$ &$-$  &$65.40\pm{0.52}$ &$0.02380\pm{0.00012}$  \\
$J$CDM$^{r_{d}}$(without $z=0.51$)  &$0.3734\pm{0.0074}$ &$-$  &$64.71\pm{0.51}$ &$0.02369\pm{0.00011}$
\\

$J$CDM  &$0.3419\pm{0.0065}$ &$-$  &$66.95\pm{0.51}$ &$0.02378\pm{0.00012}$  \\
$J$CDM(without $z=0.51$)  &$0.3563^{+0.0070}_{-0.0062}$ &$-$  &$65.87^{+0.46}_{-0.53}$ &$0.02370\pm{0.00013}$
\\

\noalign{\smallskip}\hline\noalign{\smallskip}
{\bf Non-flat} \\

$\Lambda$CDM$^{r_{d}}$   &$0.3043\pm{0.0065}$ &$0.0027\pm{0.0017}$  &$68.81\pm{0.44}$ &$0.02257\pm{0.00014}$   \\

$\Lambda$CDM$^{r_{d}}$(without $z=0.51$)  &$0.3045\pm{0.0069}$ &$0.0021\pm{0.0017}$  &$68.64\pm{0.49}$ &$0.02255\pm{0.00015}$  \\

$\Lambda$CDM   &$0.3038\pm{0.0069}$ &$0.0031\pm{0.0032}$  &$68.90\pm{0.53}$ &$0.02256\pm{0.00014}$   \\

$\Lambda$CDM(without $z=0.51$)  &$0.3035^{+0.0056}_{-0.0070}$ &$0.0024\pm{0.0032}$  &$68.80\pm{0.56}$ &$0.02258\pm{0.00014}$  \\

$J$CDM$^{r_{d}}$  &$0.3616\pm{0.0076}$ &$0.0046\pm{0.0017}$ &$66.35\pm{0.54} $&$0.02402^{+0.00011}_{-0.00013}$ \\

$J$CDM$^{r_{d}}$(without $z=0.51$)  &$0.3671\pm{0.0077}$ &$0.0010\pm{0.0018}$ &$65.42\pm{0.60} $&$0.02394^{+0.00012}_{-0.00011}$
\\

$J$CDM  &$0.3477\pm{0.0074}$ &$0.0154\pm{0.0027}$ &$69.13\pm{0.56} $&$0.02427\pm{0.00013}$ \\

$J$CDM(without $z=0.51$)  &$0.3562\pm{0.0078}$ &$0.0096^{+0.0031}_{-0.0027}$ &$67.51\pm{0.63} $&$0.02408\pm{0.00013}$
\\


\end{tabular}
\end{ruledtabular}
\end{center}
\label{table:cons_every}
\end{table*}

\begin{figure*}[htb]
\centering
\begin{minipage}[t]{0.48\textwidth}
\centering
\includegraphics[width=\linewidth]{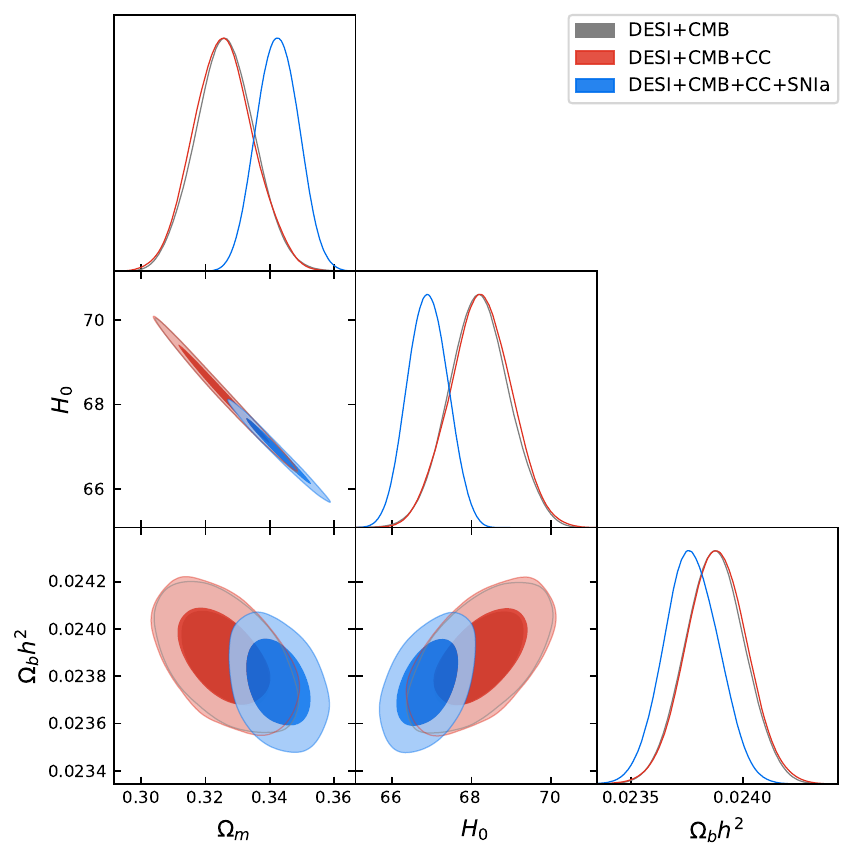}

\label{fig:flat_no_rd}
\end{minipage}
\hfill
\begin{minipage}[t]{0.48\textwidth}
\centering
\includegraphics[width=\linewidth]{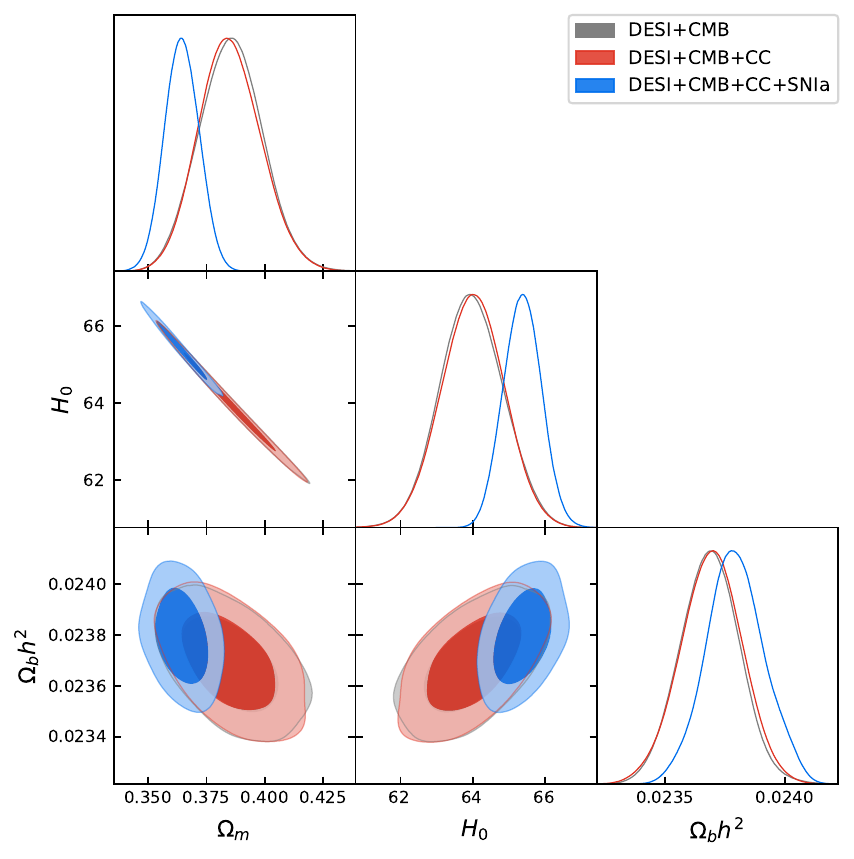}

\end{minipage}
\caption{One-dimensional marginalized probability distributions and two-dimensional confidence contour plots for the flat $J$CDM cosmological model, 
derived from the DESI+CMB, DESI+CMB+CC, and DESI+CMB+CC+SNIa datasets. Plots are presented for two acoustic horizon scale $r_{d}$ configurations: 
one defined in Eq.~(\ref{eq:rd}) (referred to as the ``unfixed” case, left panel) and the CMB derived value $r_{d} = 147.09\pm{0.29} ~\mathrm{Mpc}$ 
(referred to as the``fixed” case, right panel). Here, $H_0$ is given in units of $\rm km~s^{-1}~Mpc^{-1}$.}
\label{fig:flat}
\end{figure*}
\begin{figure*}[htb]
\centering
\begin{minipage}[t]{0.48\textwidth}
\centering
\includegraphics[width=\linewidth]{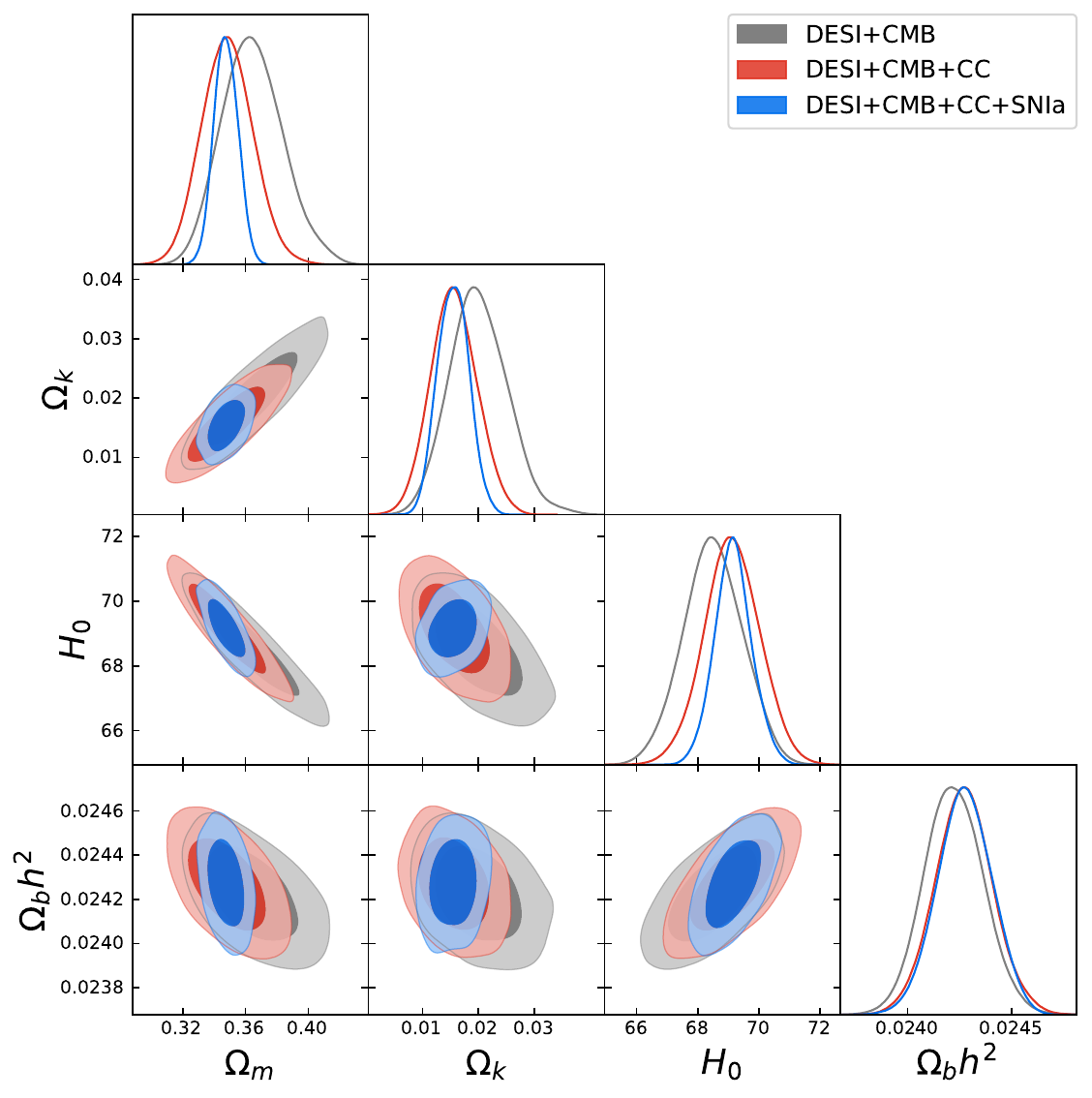}

\end{minipage}
\hfill
\begin{minipage}[t]{0.48\textwidth}
\centering
\includegraphics[width=\linewidth]{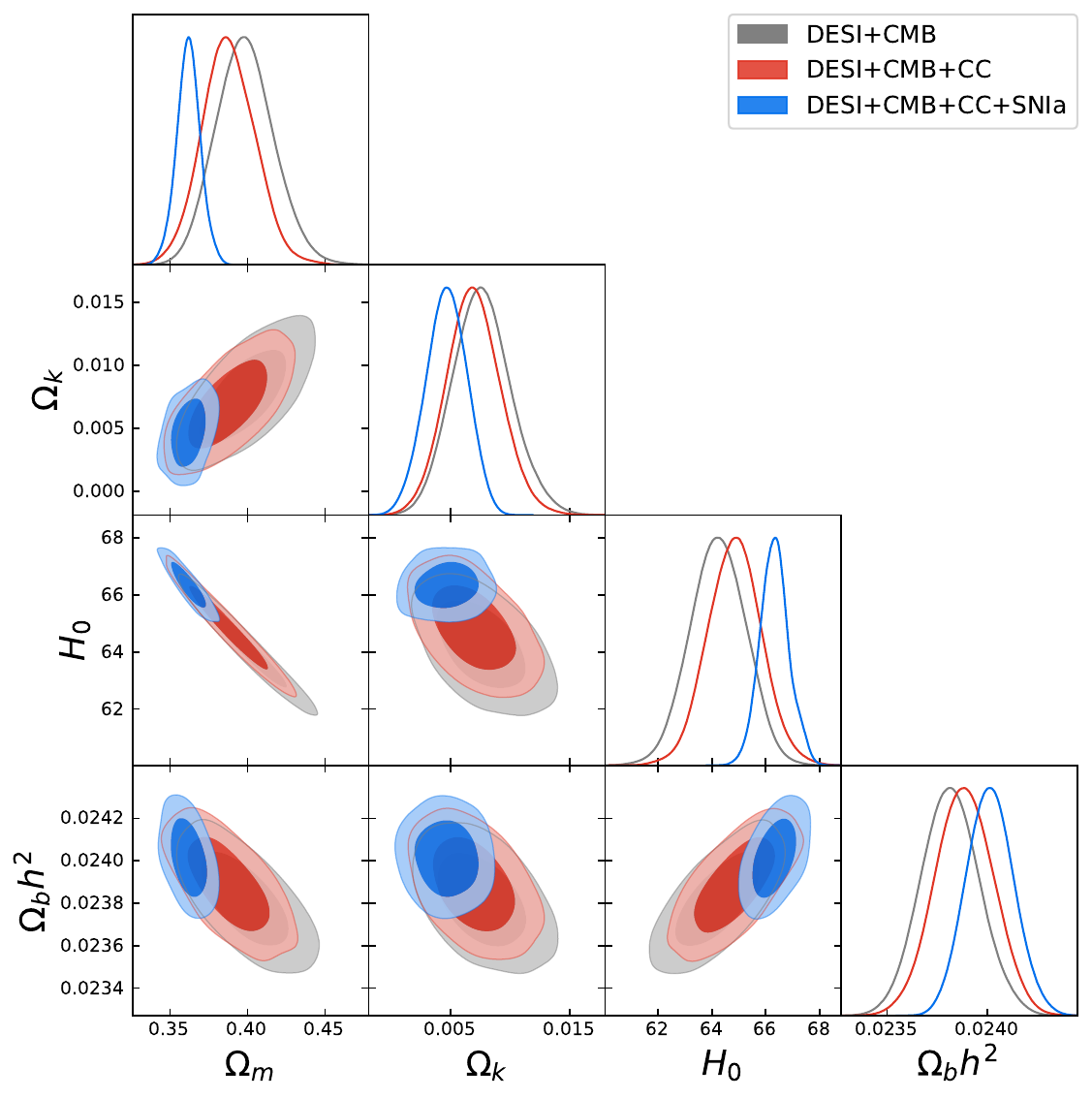}

\end{minipage}
\caption{One-dimensional marginalized probability distributions and two-dimensional confidence contour plots for the non-flat $J$CDM cosmological model, 
derived from the DESI+CMB, DESI+CMB+CC, and DESI+CMB+CC+SNIa datasets. Plots are presented for two acoustic horizon scale $r_{d}$ configurations: 
one defined in Eq.~(\ref{eq:rd}) (referred to as the ``unfixed” case, left panel) and the CMB derived value $r_{d} = 147.09\pm{0.29} ~\mathrm{Mpc}$ 
(referred to as the``fixed” case, right panel). Here, $H_0$ is given in units of $\rm km~s^{-1}~Mpc^{-1}$.}
\label{fig:non-flat}
\end{figure*}

We obtained the best-fitting parameters and their associated uncertainties using the aforementioned method, with the detailed results presented in Tables \ref{table:cons_no_rd} and \ref{table:cons_rd}. Figures \ref{fig:flat} and \ref{fig:non-flat} show the one-dimensional marginalized probability distributions and two-dimensional confidence contours for the $J$CDM model in flat and non-flat universe, respectively. These results 
are derived using two forms of $r_{d}$: one from Eq.~(\ref{eq:rd}) (left panels) and the CMB derived prior $r_{d} = 147.09\pm {0.26}\,\text{Mpc}$ (right panels), 
and are combined with dataset combinations: DESI+CMB, DESI+CMB+CC, and DESI+CMB+CC+SNIa.

The results of~\cite{van_Putten_2025} indicate that the $J$CDM model, through its dynamic dark energy mechanism, elevates the theoretically predicted \(H_0\) to approximately 74.9 $\pm 2.6$ \(\mathrm{km~s^{-1}Mpc^{-1}}\). This prediction has been validated through comparison with the local distance ladder (LDL) (see Table \ref{table:cons_every})~\cite{van_Putten_2015,VANPUTTEN2021136737}. These results are in excellent agreement with the measurements reported by the SH0ES collaboration, suggesting that the model may help alleviate the Hubble tension. In comparison, under the condition that $r_d$ is defined by Eq. (\ref{eq:rd}), our results indicate that observational constraints on \(H_0\) are $66.95\pm{0.51}~\rm km~s^{-1}Mpc^{-1}$ for a flat universe and $69.13 \pm {0.56}$ \(\mathrm{km~s^{-1}Mpc^{-1}}\) for a non-flat universe. Furthermore, with a CMB derived prior of $r_d$, the observational constraint on the Hubble constant $H_0$ is $65.40 \pm{0.52}\,\mathrm{km\,s^{-1}\,Mpc^{-1}}$ in a flat universe and $66.35 \pm 0.54\,\mathrm{km\,s^{-1}\,Mpc^{-1}}$ in a non-flat universe. These values are not only significantly lower than both the model's theoretical prediction and its observational constraints derived from the LDL, but also substantially below the direct measurements obtained in the local universe, and thus fails to effectively resolve the Hubble tension. Notably, the value of $H_0$ derived from this study is significantly lower than the value reported in~\cite{van_Putten_2025},
where $H_0$ is found to satisfy the predicted scaling relation $H_{0}=\sqrt{6/5}H^{\Lambda}_0$ following fits to local distance ladder data. 
\subsection{Impacts of acoustic horizon scale $r_d$ forms on $J$CDM model constraints}
As mentioned above, we also investigate the impact of $r_d$ forms on the final constraints for two cases: 
$r_d$ derived from Eq.~(\ref{eq:rd}) (hereafter referred to as the ``unfixed" case) and $r_d$ adopted with a CMB prior (hereafter referred to as the ``fixed" case). The corresponding results are presented in Tabs. \ref{table:cons_no_rd}, \ref{table:cons_rd}, and \ref{table:cons_every}.

In a flat universe, $H_{0}$ decreases from $66.95 \pm 0.51\ \mathrm{km\ s^{-1}\ Mpc^{-1}}$ for the unfixed $r_{d}$ case to $65.40 \pm 0.52\ \mathrm{km\ s^{-1}\ Mpc^{-1}}$ for the fixed $r_d$ case, with a $\sim 2.1\sigma$ deviation between the two values. This trend becomes even more pronounced in a non-flat universe, where $H_0$ declines from $69.13 \pm 0.56\ \mathrm{km\ s^{-1}\ Mpc^{-1}}$ to $66.35 \pm 0.54\ \mathrm{km\ s^{-1}\ Mpc^{-1}}$, representing a $3.6\sigma$ discrepancy. Such significant decline indicates that the $J$CDM model exhibits high sensitivity to the prior specification of $r_{d}$.

In contrast to the trend observed in $H_{0}$, $\Omega_m$ exhibits a systematic increase when $r_{d}$ is fixed. In a flat universe, $\Omega_m$ rises from $0.3419 \pm 0.0065$ to $0.3640 \pm 0.0073$, corresponding to a $2.3\sigma$ difference. 
In a non-flat universe, $\Omega_m$ increases from $0.3477 \pm 0.0074$ to $0.3616 \pm 0.0076$, with a $1.3\sigma$ shift. The curvature parameter $\Omega_{k}$ shows the most significant change in a non-flat universe, decreasing from $0.0154 \pm 0.0027$ to $0.0046 \pm 0.0017$, 
representing a $3.4\sigma$ difference. This behavior contrasts with the $\Lambda$CDM model, in which $\Omega_k$ does not exhibit significant 
variation between the fixed and unfixed $r_d$ cases.

\subsection{Detailed comparison of results for the unfixed $r_d$ case}

In a non-flat universe with the unfixed $r_{d}$, $J$CDM exhibits a strong preference for positive curvature ($\Omega_k = 0.0154 \pm 0.0027$), deviating from spatial flatness by more than $5\sigma$, a preference that is significantly more pronounced than that of $\Lambda$CDM ($\Omega_k = 0.0031 \pm 0.0032$). 
The Hubble constant in this scenario is $H_0 = 69.13 \pm 0.56\ \mathrm{km\ s^{-1}\ Mpc^{-1}}$, which is slightly larger than the $\Lambda$CDM value of $68.90 \pm 0.53\ \mathrm{km\ s^{-1}\ Mpc^{-1}}$. With an unfixed $r_{d}$, $J$CDM consistently yields elevated matter density estimates: $\Omega_m = 0.3419 \pm 0.0065$ in the flat universe and $0.3477 \pm 0.0074$ in the non-flat universe, both substantially higher than the corresponding $\Lambda$CDM values of $0.3001 \pm 0.0255$ and $0.3038 \pm 0.0069$, respectively. This suggests that the enhanced matter density (unfixed $r_d$ case) is an intrinsic property of the $J$CDM model's dynamic dark energy formulation, rather than an artifact of specific data treatment or analysis choices.

The comparison of different dataset combinations presented in Tab.~\ref{table:cons_no_rd} further highlights the evolution of $J$CDM's parameter preferences. As additional observational datasets are progressively incorporated, starting from DESI+CMB, then adding CC, and finally including SNIa, $\Omega_{m}$ increases systematically while $H_0$ decreases correspondingly. This trend indicates that SNIa data play a pivotal role in steering the $J$CDM model toward solutions characterized by higher matter density.

\begin{figure*}[htb]
\centering
\begin{minipage}[t]{0.48\textwidth}
\centering
\includegraphics[width=\linewidth]{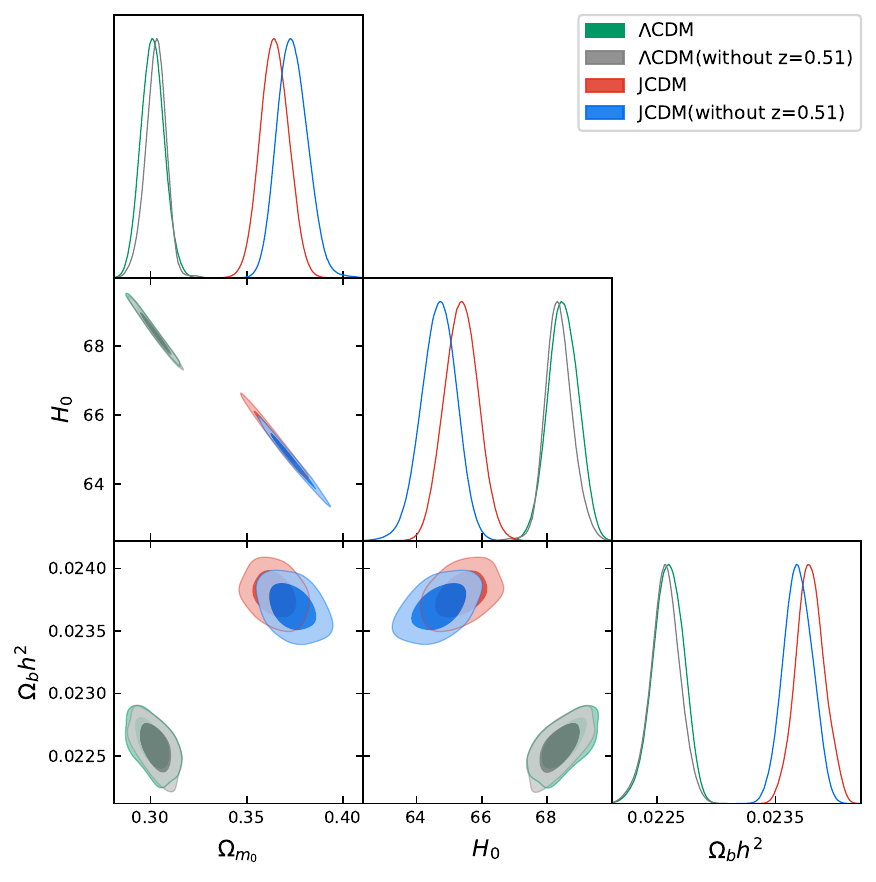}

\end{minipage}
\hfill
\begin{minipage}[t]{0.48\textwidth}
\centering
\includegraphics[width=\linewidth]{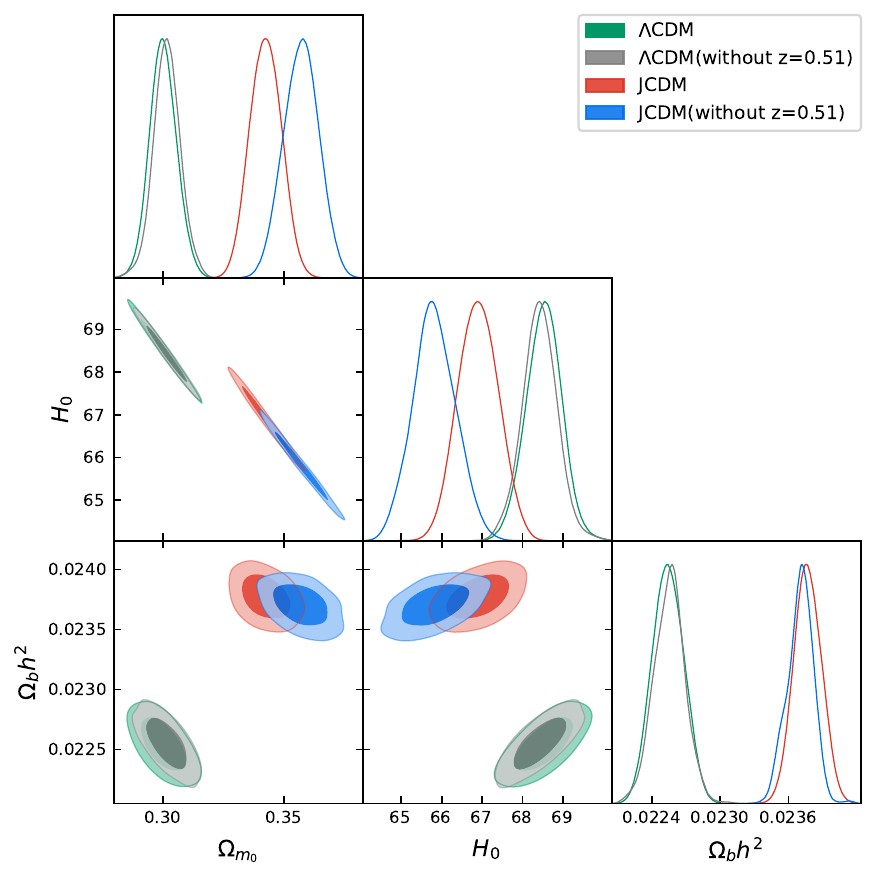}

\end{minipage}
\caption{One-dimensional marginalized probability distributions and two-dimensional confidence contour plots for the flat $\Lambda$CDM and 
$J$CDM cosmological models, 
derived from full combined datasets DESI+CMB+CC+SNIa. Plots are presented for two acoustic horizon scale $r_{d}$ configurations: 
one defined in Eq.~(\ref{eq:rd}) (referred to as the ``unfixed” case, left panel) and the CMB derived value $r_{d} = 147.09\pm{0.29} ~\mathrm{Mpc}$ 
(referred to as the``fixed” case, right panel). For comparison, results with and without the DESI BAO data at $z_{\rm eff}=0.51$ (shown as $z=0.51$) are also plotted. 
Here, $H_0$ is given in units of $\rm km~s^{-1}~Mpc^{-1}$.}
\label{fig:3}
\end{figure*}
\begin{figure*}[htb]
\centering
\begin{minipage}[t]{0.48\textwidth}
\centering
\includegraphics[width=\linewidth]{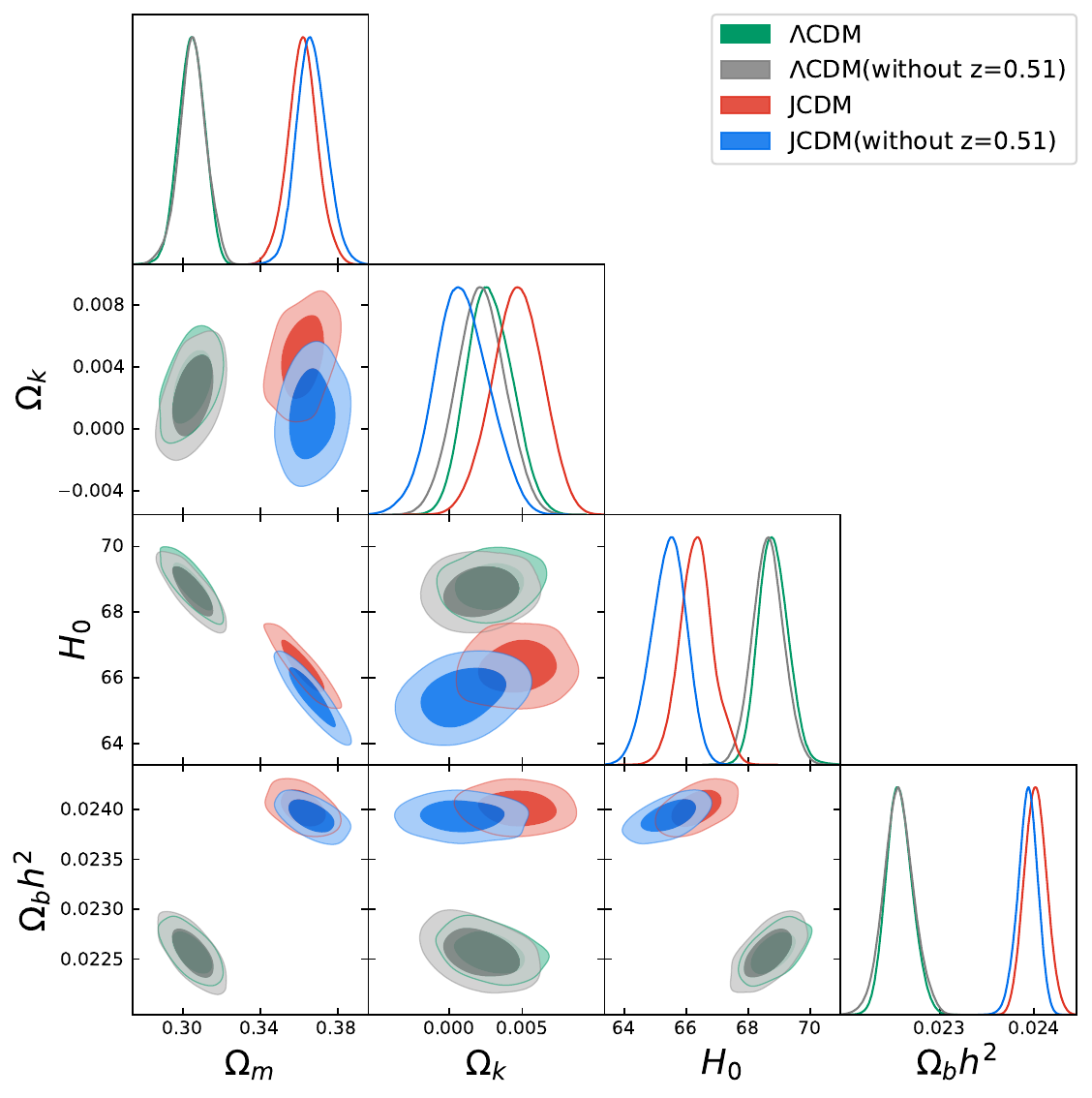}

\end{minipage}
\hfill
\begin{minipage}[t]{0.48\textwidth}
\centering
\includegraphics[width=\linewidth]{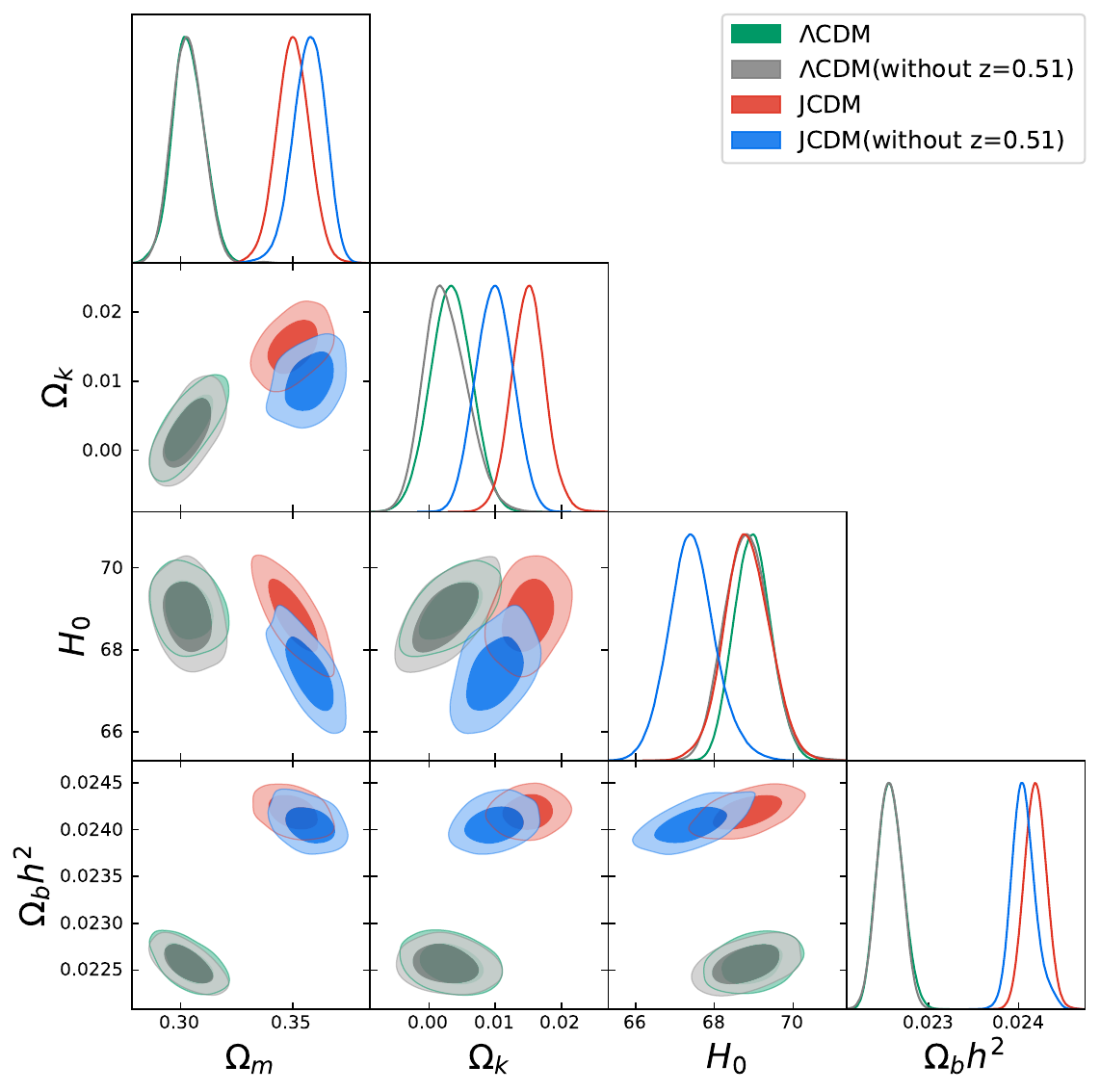}

\end{minipage}
\caption{One-dimensional marginalized probability distributions and two-dimensional confidence contour plots for the non-flat $\Lambda$CDM and 
$J$CDM cosmological models, 
derived from full combined datasets DESI+CMB+CC+SNIa. Plots are presented for two acoustic horizon scale $r_{d}$ configurations: 
one defined in Eq.~(\ref{eq:rd}) (referred to as the ``unfixed” case, left panel) and the CMB derived value $r_{d} = 147.09\pm{0.29} ~\mathrm{Mpc}$ 
(referred to as the``fixed” case, right panel). For comparison, results with and without the DESI BAO data at $z_{\rm eff}=0.51$ (shown as $z=0.51$) are also plotted. 
Here, $H_0$ is given in units of $\rm km~s^{-1}~Mpc^{-1}$.}
\label{fig:4}
\end{figure*}

\subsection{Detailed comparison of results for the fixed $r_d$ case}
Within the framework of fixed $r_{d}$, the differences between the $J$CDM and $\Lambda$CDM models become most prominent, with both models constrained under identical early universe priors. In a flat universe, $J$CDM yields $\Omega_m = 0.3640 \pm 0.0073$, significantly higher than the $\Lambda$CDM value of $0.2998^{+0.0045}_{-0.0056}$, corresponding to an approximately $6.6\sigma$ discrepancy; this trend persists in the non-flat case, where $J$CDM gives $\Omega_m = 0.3616 \pm 0.0076$, compared to the value of $\Lambda$CDM model $\Omega_m = 0.3043 \pm 0.0065$. The systematic overestimation of matter density suggests that 
$J$CDM's dynamic dark energy mechanism may require additional matter components to remain consistent with the observed cosmic expansion history. In contrast, in a flat universe, $J$CDM predicts $H_0 = 65.40 \pm 0.52\ \mathrm{km\ s^{-1}\ Mpc^{-1}}$, which is significantly lower than the $\Lambda$CDM result of $68.58^{+0.42}_{-0.36}\ \mathrm{km\ s^{-1}\ Mpc^{-1}}$ derived from CMB data, and lies well below the local distance ladder measurement ($73.04 \pm 1.04\ \mathrm{km\ s^{-1}\ Mpc^{-1}}$). Considering the original Hubble tension, this result reveals a tension between $J$CDM and early-universe physics as inferred from the CMB, now in the direction opposite to that of the original Hubble tension between local distance ladder and CMB measurements in the $\Lambda$CDM model.

Furthermore, in the non-flat geometry scenario with a fixed $r_d$, both models favor a nearly flat universe; however, $J$CDM predicts a marginally higher curvature, with $\Omega_k = 0.0046 \pm 0.0017$, compared to the value of $\Lambda$CDM model $\Omega_k = 0.0027 \pm 0.0017$. This residual difference suggests that even under a strong $r_d$ prior, the dynamic dark energy mechanism in the $J$CDM model continues to exert a subtle yet detectable influence on the universe's geometric properties.


\subsection{Impacts of the DESI BAO data point at $z_{\rm eff}={0.51}$ on the final constraints}
The Dark Energy Spectroscopic Instrument (DESI) has reported an anomalously high measurement of the matter density at an effective redshift of $z_{\rm eff} = 0.51$ in its luminous red galaxy (LRG) sample. This single data point exhibits significant tension with results from other cosmological probes. In this section, we present a systematic analysis of its impact on the parameter constraints of the standard $\Lambda$CDM model and the alternative $J$CDM model. Table~\ref{table:cons_every} summarizes the model parameter constraints derived from the combined dataset. Figures~\ref{fig:3} and~\ref{fig:4} present the one-dimensional marginalized probability distributions and two-dimensional confidence contours for the standard $\Lambda$CDM cosmological model and $J$CDM model in both flat and non-flat universe. The left (right) panels corresponds to the case with a fixed (unfixed) $r_{d}$.

For a flat universe and fixed $r_d$, with and without the data point at $z_{\rm eff}=0.51$ has a slight impact on the parameters of the $\Lambda$CDM model. The Hubble constant $H_0$ shifts from $68.58^{+0.42}_{-0.36}$ to $68.42 \pm 0.39\ \mathrm{km\ s^{-1}\ Mpc^{-1}}$, while the matter density parameter $\Omega_m$ changes from $0.2998^{+0.0045}_{-0.0066}$ to $0.3019 \pm 0.0051$. For the unfixed $r_d$ case, the resulting parameter variations remain minimal. The Hubble constant decreases slightly from $68.53 \pm 0.43$ to $68.43 \pm 0.44\ \mathrm{km\ s^{-1}\ Mpc^{-1}}$, and the matter density parameter increases marginally from $0.3001 \pm 0.0255$ to $0.3016 \pm 0.0055$. Notably, the exclusion of this data point leads to a significant reduction in the uncertainty of $\Omega_m$, indicating that the $z_{\rm eff}=0.51$ measurement constitutes one of the primary sources of total uncertainty in matter density estimation within the $\Lambda$CDM framework.

In the non-flat universe scenario, the $\Lambda$CDM model continues to exhibit strong stability. With a fixed $r_d$, the spatial curvature parameter $\Omega_k$ evolves from $0.0027 \pm 0.0017$ (with $z_{\rm eff}=0.51$) to $0.0021 \pm 0.0017$ (without $z_{\rm eff}=0.51$), remaining consistent with a nearly flat geometry; the Hubble constant decreases modestly from $68.81 \pm 0.44$ to $68.64 \pm 0.49\ \mathrm{km\ s^{-1}\ Mpc^{-1}}$. The small magnitude of these changes further underscores the robustness of the $\Lambda$CDM model against outlier data points. In contrast to the $\Lambda$CDM model, the $J$CDM model exhibits pronounced sensitivity to the $z_{\rm eff}=0.51$ data point. In a flat universe with fixed $r_d$, removing this observation results in a decrease in $H_0$ from $65.40 \pm 0.52$ to $64.71 \pm 0.51\ \mathrm{km\ s^{-1}\ Mpc^{-1}}$, accompanied by an increase in $\Omega_m$ from $0.3640 \pm 0.0073$ to $0.3734 \pm 0.0074$. 

For the flat $J$CDM universe with an unfixed $r_d$, the Hubble constant $H_0$ decreases from $66.95 \pm 0.51$ (with $z_{\rm eff}=0.51$) to $65.87^{+0.46}_{-0.53}\ \mathrm{km\ s^{-1}\ Mpc^{-1}}$ (without $z_{\rm eff}=0.51$), while $\Omega_m$ increases from $0.3419 \pm 0.0065$ to $0.3563^{+0.0070}_{-0.0062}$. In the non-flat $J$CDM universe, the effects are even more pronounced. With an unfixed $r_d$, exclusion of the $z_{\rm eff}=0.51$ data point causes the spatial curvature parameter $\Omega_k$ to decline from $0.0154 \pm 0.0027$ to $0.0096^{+0.0031}_{-0.0027}$, and $H_0$ decreases from $69.13 \pm 0.56$ to $67.51 \pm 0.63\ \mathrm{km\ s^{-1}\ Mpc^{-1}}$. This behavior reveals a key characteristic of the $J$CDM framework: under a unfixed $r_d$, the model relies on positive spatial curvature to alleviate tensions between the $z_{\rm eff}=0.51$ measurement and other cosmological constraints.  

Moreover, the baryon density parameter $\Omega_b h^2$ in $J$CDM also responds to the removal of the $z_{\rm eff}=0.51$ data point. In the flat universe and fixed $r_d$ scenario, it decreases from $0.02380 \pm 0.00012$ to $0.02369 \pm 0.00011$. Nonetheless, its value remains systematically higher than that obtained under the $\Lambda$CDM model, highlighting a persistent discrepancy across modeling assumptions.

As noted earlier, DESI analyses may contain internal tensions, plausibly linked to the use of a fiducial $\Lambda$CDM cosmology in BAO reconstruction to correct for cosmic evolution. The late-time expansion history is independently probed over a comparable redshift range by the local distance ladder and by cosmic chronometers. As shown in Tab.~\ref{table:cons_every}, the $\Lambda$CDM results derived from DESI exhibit remarkably close agreement with those from Planck constraints. This level of concordance is closer than might be expected for nominally independent datasets probing very different epochs. At the same time, $\Lambda$CDM fits to DESI data do not reproduce the large tension reported by the local distance ladder. Given that BAO reconstruction in DESI relies on a fiducial $\Lambda$CDM cosmology to correct for cosmic evolution, this raises the possibility that reconstruction-related model assumptions may reduce the diagnostic power of DESI BAO measurements when used to test alternatives to $\Lambda$CDM. Therefore, in their current form, DESI BAO data may not be optimally suited for robust, model-agnostic tests of alternative cosmologies.

\section{Conclusions}
\label{sec:conclusions}

We have investigated the cosmological constraints on the big bang quantum cosmology model 
(referred to as the $J$CDM model) using observational data from DESI BAO, CMB, CC, and SNIa observations. Our findings reveal that, when 
adopting the value of $r_d$ derived from the combination of cosmological parameters (Eq.~(\ref{eq:rd})), the matter density parameter $\Omega_m$ 
is constrained to \(0.3419\pm{0.0065}\), and the Hubble constant $H_0$ is constrained to \(66.95\pm{0.51}\, \mathrm{km~s^{-1}Mpc^{-1}}\) for the flat universe 
scenario. For the non-flat universe, $\Omega_m$ is constrained to
\(0.3477\pm{0.0074}\), $H_0$ to \(69.13\pm{0.56}\, \mathrm{km~s^{-1}Mpc^{-1}}\), and the curvature parameter \(\Omega_k\) to $0.0154\pm{0.0027}$, 
suggesting the possibility of a positively curved universe. Note that in the case of a flat universe, the value of the matter density parameter $\Omega_m$ is greater than that obtained from Planck 2018 data, whereas the Hubble constant $H_0$ is smaller. 
For a non-flat universe, both $\Omega_m$ and $H_0$ exhibit larger values compared to those derived from Planck 2018. 

We have also investigated the impact of adopting $r_d$ in different forms on the final constraints: one derived from Eq.~(\ref{eq:rd}) (unfixed), 
and the other from a CMB derived prior (fixed). 
For both cases of flat and non-flat universe, the final constraints on various parameters exhibit significant differences between the unfixed and 
fixed $r_d$ cases. Additionally, we explored the influence of the DESI BAO data at $z_{\rm eff}=0.51$ on the final constraints. Consistent with the aforementioned results, in both flat and non-flat universe, the final constraints on different parameters also show notable changes when comparing the unfixed and fixed $r_d$ scenarios. Given the characteristics of the DESI BAO data, they may not yet be in a form that is ideally suited for robust, model-agnostic tests of alternative cosmologies. 

Based on our findings, the $J$CDM model displays distinct characteristics when confronted with current observational data. While the $J$CDM model, which introduces dark energy from first principles, has previously been shown to account for the high values from local distance ladder observations in a physically motivated and non-fine-tuned way, our results indicate that $J$CDM does not extend consistently to CMB measurements without tension. 
Therefore, it should be noted that $\Lambda$CDM and $J$CDM each succeed in complementary domains: $\Lambda$CDM aligns with early-universe constraints, 
while $J$CDM describes late-time cosmology. $J$CDM, while providing an excellent fit to late-time observations
such as the local distance ladder, struggles to simultaneously
accommodate early-universe constraints from Planck. This highlights that both models are partially successful, and further observational tests 
are needed to clarify the intrinsic behavior of $J$CDM.

\section{Acknowledgements}
We thank Yan Gong for useful comments and suggestions. This work is supported by the Shandong Provincial Natural Science Foundation 
(Grant Nos.ZR2025MS16,ZR2025MS47, ZR2025QC25).
\
\newpage
\bibliographystyle{apsrev4-1}
\bibliography{ref}
\end{document}